\documentclass[a4paper,twocolumn,10pt]{article}

\usepackage{spects}
\usepackage{graphicx}
\usepackage{amsmath}

\begin{document}

%
%
\title{An
intelligent floor field cellular automata model for pedestrian
dynamics}
\author{Ekaterina Kirik, Tat'yana Yurgel'yan, Dmitriy Krouglov
\\
Institute of Computational Modelling of Siberian Branch of Russian Academy of Sciences, Siberian Federal University\\
50 Akademgorodok, Krasnoyarsk, Russia, 660036; 79 Svobodniy av., Krasnoyarsk, Russia, 660041\\
kirik@icm.krasn.ru}

\maketitle



\begin{abstract}
\noindent A stochastic cellular automata (CA) model for pedestrian
dynamics is presented.  Our goal is to simulate different types of
pedestrian movement, from regular to panic. But here we emphasize
regular situations which imply that pedestrians analyze
environment and choose their route more carefully. And transition
probabilities have to depict such effect. The potentials of floor
fields and environment analysis are combined in the model
obtained. People patience is included in the model. This  makes
simulation of pedestrians movement more realistic. Some simulation
results are presented and comparison with basic FF-model is made.
\end{abstract}

%
%
\newcommand{\blabla}{body text body text body text body text body text body text}
\newcommand{\blachap}{Body text \blabla \blabla \blabla \blabla \blabla \blabla \par}

\section{Introduction}
Modelling of pedestrian dynamics is actual problem at present
days. Different approaches from  the social force
model~(\cite{HelbingOV} and references therein) based on
differential equations to stochastic CA models~(\cite{ExtFFCAMod,
MalStep, Yamamoto} and references therein) are developed. They
reproduce many collective properties including lane formation,
oscillations of the direction at bottlenecks, the so-called
``faster-is-slower'' effect. These are an important and remarkable
basis  for pedestrian modelling. But there are still things to be
done in order to reproduce individual pedestrian behavior more
realistic and carefully.

 The model presented takes its inspiration from stochastic floor
field~(FF) CA model~\cite{ExtFFCAMod}. Here a static field is a
map that pedestrian may use to orient in the space. Dynamic field
is used to model herding behavior in panic situations.

It's known that regular situations imply that pedestrians analyze
environment and choose their route more carefully
(see~\cite{HelbingOV} and reference therein). Pedestrians keep a
certain distance from other people and obstacles. The more hurried
a pedestrian is and more tight crowd is  this distance is smaller.
We adopted a mathematical formalization of these points from
~\cite{MalStep}.

Pedestrians minimize efforts to reach their destinations:  feel
strong aversion to taking detours or moving opposite to their
desired direction. However, people normally choose the fastest
rout but not the shortest. This means that opportunity to wait (to
stay at present place) has to be realized in the model.
(Models~\cite{ MalStep, ExtFFCAMod} (and other CA models) imply
that people can stay at present place if there is no space to move
only.) We realize this point (people patience) in the algorithm.

As well as it's necessary to take into account that some effects
are more reside for certain regions. For instance, clogging
situations are more pronounced in the nearest to an exit areas.
This means that spatial adaptivity of correspondent model
parameters to be introduced in the model. All these changes and
additions extend basis FF model towards emotional aspect and
improve and make flexible decision making process. By this reason
model obtained was named as {\it Intelligent FF model}.

\section{Intelligent floor field  model}
\subsection{Space structure}
As usual for CA models the space (plane) is sampled into cells
$40cm \times 40cm$ (it's an average space occupied by a pedestrian
in a dense crowd ~\cite{SimEvacuatFFM}) which can either be empty
or occupied by one pedestrian (particle) only.

 The von Neumann neighborhood is used. It
implies that each particle can move to one of four its
next-neighbor cells $(i,j)$ or to stay at the present cell at each
discrete time step $\,t \rightarrow t+1$, e.i., $v_{max}=1$.
(Empirically the average velocity of a pedestrian is about $1.3
m/s$. So real time corresponding to one time step in the model is
about $0.3 s$. ) Such movement is in accordance with certain
transition probabilities that are explained below.

\subsection{Floor fields}
Static ($S$) and dynamic ($D$) floor fields are introduced and
discussed in~\cite{SimPedDynFFM, SimEvacuatFFM, ExtFFCAMod}. For
each cell $(i,j)$ values of $S_{ij}$ and $D_{ij}$ are given.

\textit{Static floor field} $S$ describes the shortest distance to
an exit (or other destination point that depends on a task). It
doesn't evolve with time and isn't changed by the presence of the
particles. The value of $S_{ij}$ is set inversely proportional to
the distance from the cell $(i,j)$ to the exit. One can consider
$S$ as a map that pedestrian can use to move to the target point,
e.g., exit.

\textit{Dynamic floor field} $D$ is a virtual trace left by the
pedestrians similar to the pheromone in chemotaxis. It is used to
model a " long-ranged'' attractive interactions between the
pedestrians, e.g., herding behavior that is observed in panic
situations. Dynamic floor field is time dependent. In each time
step each $D_{ij}$ decays with probability $\delta$ and diffuses
with probability $\alpha \in [0,1]$ to one of its four neighboring
cells.  Decay and diffusion lead to broadening, dilution, and
finally vanishing of the trace. At $t=0$ for all cells $D_{ij}=0$.
$D_{ij} \rightarrow D_{ij}+1$ just after particle left cell
$(i,j)$.

Field $S$ and $D$ works in such a way that transition probability
increases in the direction of higher fields $S$ and $D$, i.e.
motion in such direction is more feasible.

\subsection{Environment analysis}
To make next step  pedestrian observes surroundings. Let $r>0$ be
maximum distance (in cells) at which pedestrian can see around, by
$$f_{ij}= \left\{%
\begin{array}{ll}
    1, & \hbox{cell $(i,j)$ is occupied by a pedestrian;} \\
    0, & \hbox{cell $(i,j)$ is empty;} \\
\end{array}%
\right. $$ denote the occupation number.

If desired direction (cell) is occupied to minimize efforts to
reach destination  or to realize the fasters rout pedestrian  has
to a have an opportunity to stay at present cell or move to one of
the rest unoccupied nearest cells.

Keeping apart from other people and obstacles can be simulated by
decreasing probability for such direction (let $\alpha$ be a name
of the direction). For present position of some pedestrian let
cell $(i,j)$ be a next-neighbor in the direction $\alpha$.  For
cell $(i,j)$ we will calculate a term~\cite{MalStep}
$$A_{ij}=\bigl(1-\frac{1}{r}(\sum\limits_{m=i \,(l=j)}^{i\pm r^{\star}_{ij}(j\pm r^{\star}_{ij})}
f_{ml}+r-r^{\star}_{ij})\bigr),$$ where $r^{\star}_{ij}$ ---
distance to a nearest obstacle (wall, column, etc., but not
pedestrian) in the direction $\alpha$ starting from cell $(i,j)$
($r^{\star}_{ij}\leq r$); $\sum\limits_{m=i \,(l=j)}^{i \pm
r^{\star}_{ij}(j \pm r^{\star}_{ij})} f_{ml}$
--- number of pedestrians that are in the direction $\alpha$ starting from cell $(i,j)$
up to the nearest obstacle (sum is over lines ($m$) or over
columns ($l$) depending on direction $\alpha$), all cells behind
obstacle are considered as occupied.

\subsection{Update rules}
Update rules for CA are the following:
\begin{enumerate}
  \item For each pedestrian the transition probability $p_{ij}$ to move
  to cell $(i,j)$ one of four next-neighbors is
\begin{multline}
\label{1} p_{ij}= Norm^{-1} \bigl[
      \bigl(1-\frac{1}{r}(\sum_{m=i\,(l=j)}^{i \pm r^{\star}_{ij}(j \pm r^{\star}_{ij})} f_{ml}+r-r^{\star}_{ij})\bigr) \\
      \exp(k_S S_{ij}) \exp(k_D D_{ij}) \exp (k_I) 
\bigr],
\end{multline}
\noindent where $k_S$,$k_D$ --- sensitivity parameters; $\exp
(k_I)$ --- inertia effect: $k_I>0$ for the direction of
pedestrian's motion in the previous time step and $k_I=0$ for
other cells; normalization
$$Norm = \sum\limits_{(i,j)} A_{ij} e^{k_S S_{ij}} e^{k_D D_{ij}}
       e^{k_I},$$
where the sum is over all possible target cells.

  \item If $Norm=0$ then pedestrian stays at the present cell,
  otherwise pedestrian chooses randomly a target cell $(i,j)^{\ast}$
  based on the transition probabilities determined by
  (1).
  \item If $Norm\neq0$ and  $(1-f_{ij}^{\ast})=0$ (i.e., cell $(i,j)^{\ast}$ is occupied)
  then  pedestrian  chooses randomly a target cell once again.
  Now target cell is chosen among the following candidates: rest next-neighbors available for moving
  (i.e., $(1-f_{ij})\neq 0$) and the present
  cell. For the available next-neighbor cell $(i,j)$ new transition probability is
  $\frac{p_{i,j}}{\sum\limits_{(i,j)} p_{i,j}}$
  and pedestrian can stay at present cell with probability
  $\frac{p_{i,j}^{\ast}}{\sum\limits_{(i,j)} p_{i,j}}$ (where sum is over all available candidates at now).
   Obviously, if there is no available next-neighbors then
  particle stays at present cell.

  In contrast to~\cite{MalStep, ExtFFCAMod} this step gives an
  opportunity for pedestrians not to move and wait when preferable
  direction will free.

  \item Whenever two or more pedestrians  have the same target cell,
  the movement of all involved pedestrians is denied with
  probability $\tilde{\mu}_{ij}$, i.e. all pedestrians remain at their
  old places~\cite{ExtFFCAMod}. One of
  the candidates moves to the desired cell with the probability $1-\tilde{\mu}_{ij}$. Pedestrian that is allowed to
  move has the largest probability among all candidates (in this case probability is a measure of
  pedestrians physical strength).
  The other probabilistic method~\cite{SimPedDynFFM, MalStep, ExtFFCAMod} can be used here as well.
  \item Pedestrians that are allowed to move perform their
  motion to the target cell. $D$ at the origin cell $(i,j)$ of
  each moving particle is increased by one $D_{ij} \rightarrow
  D_{ij}+1$ and therefor can take any non-negative integer value.
\end{enumerate}

These rules are applied to all particles at the same time, i.e.,
parallel update is used.

\subsection{Model parameters}
There are several parameters in the model obtained. Their values
and physical meaning are presented below.
\begin{itemize}
  \item $r>0$ --- maximal distance at which pedestrian can feel
  the surroundings. People avoid to walk close to obstacles and
  other people. $A_{ij}=0$ if there is no free space to move, $A_{ij}=1$
  if direction is free, $0 < A_{ij} < 1$ for any other intermediate situation.
  \item $k_S \geq 0$ --- sensitivity parameter that can be
  interpreted as the knowledge of the shortest way to the destination point,
  or as a wish to move in a certain direction.
  $k_S = 0$ means that pedestrian don't use information from the field $S$.
  The higher $k_S$ is movement of the pedestrians is more directed.
  \item $k_D \geq 0$ --- sensitivity parameter that can be
  interpreted as a rate of herding behavior. It is known that people try to follow others particulary in
panic situations ~\cite{HelbingOV}. $k_D=0$ means that pedestrian
chooses a way of their own ignoring ways of others. The higher
$k_D$ is the herding behavior of the pedestrians is more
pronounced.
  \item $k_I \geq 0$ --- parameter that determines the strength of inertia which suppresses
quick changes of the direction.
  \item $0 \leq \tilde{\mu}_{ij} \leq 1$:
  $$\tilde{\mu}_{ij}=
  \begin{cases}
     \frac{S_{ij}}{\max\limits_{i,j}
  S_{ij}}\mu, & \text{if  } k_S \neq 0, \\
    \mu, & \text{otherwise},
  \end{cases}$$
where $\mu \in [0, 1]$  --- friction parameter that controls the
resolution of conflicts in clogging situations. $\tilde \mu_{ij}$
works as some kind of local pressure between the pedestrians. The
higher $\tilde \mu_{ij}$ is pedestrians are more handicapped by
others trying to reach the same target cell. Such situations are
natural and well pronounced for nearest to exit (destination
point) space. For other areas it's not typical but it's possible.
So to realize it and make simulation of individuals realistic the
coefficient $\frac{S_{ij}}{\max\limits_{i,j}
  S_{ij}}$ is introduced (in contrast with original FF model~\cite{ExtFFCAMod}).
  \item  $\delta$, $\alpha \in [0,1]$ --- these constants control diffusion and
decay of the dynamic floor field~\cite{ExtFFCAMod}. ``It reflects
the randomness of people's movement and the visible range of a
person, respectively. If the room is full of smoke, then $\delta$
takes large value due to the reduced visibility. Through diffusion
and decay the trace is broadened, diluted and vanishes after some
time.''
\end{itemize}

\section{Discussion of the model}
Model obtained is simple. For one time step there are $O(n)$
calculations if $n$ pedestrians are involved. It gives advantage
over continuous social-force model~\cite{HelbingOV} where each
time step $O(n^2)$ interaction terms have to be evaluated. The
discreteness of the model is advantage as well. It allows for a
very efficient implementation for large-scale computer
simulations.

It's shown~\cite{ ExtFFCAMod, SimPedDynFFM, SimEvacuatFFM} that
original FF model reproduces variety of collective effects:
clogging at large densities, lane formation in counterflow,
oscillation in counterflow at bottlenecks, patterns at
intersection, trail formation, ``faster-is-slower" and
``freezing-by-hearting" (in panic). All these effects are
simulated by varying of the model parameters. Model obtained saves
this opportunities. Modifications and improvements made here
mainly concern the quality of pedestrians behavior reproducing.
They allow more realistically (carefully) simulate analysis that
people accomplish while they choose a direction for moving. Idea
of parameters adaptivity makes model more flexible and closer to
real life.

At first let us consider components that determine probability
$p_{ij}$. In contrast to FF model~\cite{ExtFFCAMod} one can
distinguish two different types of terms in (\ref{1}).

Term $A_{ij}$ characterizes the physical possibility to move,
$A_{ij} \in[0,1]$. It  takes maximal value if movement conditions
in the direction are favorable. And $A_{ij}=0$ if there is no free
space to move. Term $A_{ij}$ proportionally decreases with the
advent and approaching of some obstacles (people, wall, etc.) in
the direction.

Other terms $e^{k_S S_{ij}}$, $e^{k_D D_{ij}}$, $e^{k_I}$ vary
form $1$ to $\infty$ (in general case) and characterize style of
people behavior. Minimal value of parameter ($k_S=0$, $k_D=0$ or
$k_I=0$) means that correspondent feature of behavior isn't
realized  and term doesn't affect the probability. If all three
terms are minimal then pedestrians walk free. And in this case
only term $A_{ij}$ determines the transition probability for each
next-neighbor cell $(i,j)$ in accordance with people features:
keeping apart from other people and obstacles, patience.

In FF model~\cite{SimPedDynFFM, SimEvacuatFFM, ExtFFCAMod}
pedestrians stay at present cell if there is no space to move
only. Here we give pedestrians the opportunity to wait when
preferable direction will free even if other directions are
available for moving at this time. Such behavior is reside to low
and middle densities. To realize it transition probabilities
(\ref{1}) don't include a checking if cell $(i,j)$ occupied or
not. In this case transition probabilities (\ref{1}) can be
considered as a rate of wish to go to the certain directions.
Possibility not to leave present cell is realized in step 3.

The idea of spatial adapted parameters is introduced here. It's
clear that conditions can't be equal for all people involved.
Position of pedestrian in the space may determine some features of
the behavior. Thus clogging situations are natural and well
pronounced for nearest to exit (destination point) space. For
other areas it's not typical but it's possible. To realize it the
coefficient $\frac{S_{ij}}{\max\limits_{i,j}S_{ij}}$ for clogging
parameter $\mu$ is introduced.

\section{Simulation results}
In order to test our model a regular evacuation process was
simulated. This means that $k_D=0$ in examples presented. And we
set $k_I=0$, $\mu=0$.

\subsection{One pedestrian case}
There was simulated evacuation of one person (n=1) from a room
$6.8m\times6.8m$ ($17$ cells $\times$ $17$ cells) with one exit
($0.8m$) in the middle of a wall. Recall that the space is sampled
into cells of size $40cm \times 40cm$ which can either be empty or
occupied by one pedestrian only. Static field $S$ was calculated
in accordance with ~\cite{SimEvacuatFFM}. Stating position is a
cell in a corner near wall opposite to the exit.
Pedestrian moves towards the exit with $v=v_{max}=1$. 
For such sampled space minimal value of time steps that require to
leave the room starting from initial position is $T_{min}=26$.

Different combinations of parameters $k_S$ and $r$ were
considered. Total evacuation time and trajectories were
investigated. Following table contains results over 500
experiments.

\begin{table}[!h]
\center \caption{Modes $T_{mo}$ of total evacuation time
distributions in one pedestrian case.} \label{Tmo} \fbox{
\begin{tabular}{lrr} $k_S$ & \multicolumn{1}{c}{$r$} &
\multicolumn{1}{c}{$T_{mo}$}\\[4pt]
1 & 1 & 45 \\
1 & 8 & 40 \\
1 & 17& 35 \\
\end{tabular}}
\fbox{
\begin{tabular}{lrr} $k_S$ & \multicolumn{1}{c}{$r$} &
\multicolumn{1}{c}{$T_{mo}$}\\[4pt]
2 & 1 & 29 \\
2 & 8 & 29 \\
2 & 17& 27 \\
\end{tabular}}
\fbox{
\begin{tabular}{lrr} $k_S$ & \multicolumn{1}{c}{$r$} &
\multicolumn{1}{c}{$T_{mo}$}\\[4pt]
4 & 1 & 26 \\
4 & 8 & 26 \\
4 & 17& 26 \\
\end{tabular}}
\end{table}

\begin{figure}[!h]
\begin{center}
\includegraphics[scale=0.453]{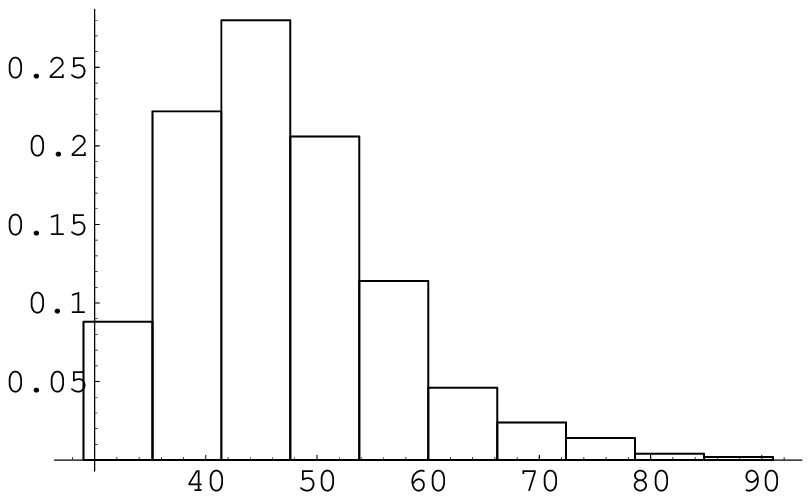}
\includegraphics[scale=0.453]{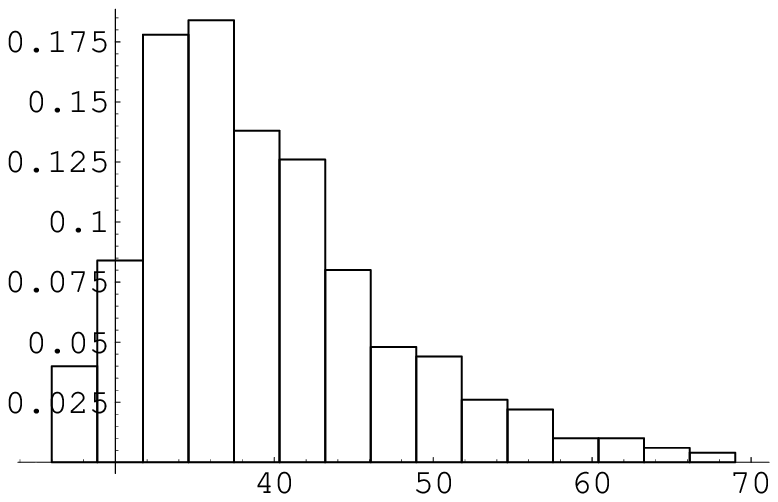}\\
\hspace{0.5cm}\parbox[t]{3.5cm}{a) $k_S=1$, $r=1$.}
\hspace{0.5cm}\parbox[t]{3.5cm}{b) $k_S=1$, $r=17$.}\\
\includegraphics[scale=0.4533]{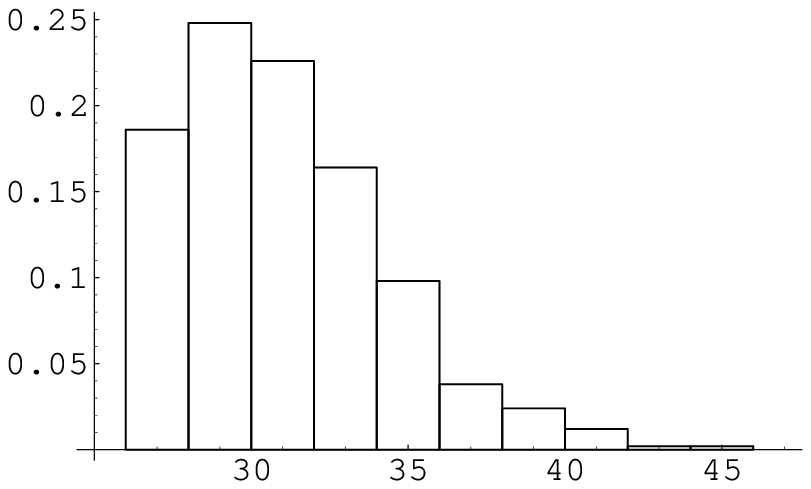}
\includegraphics[scale=0.4533]{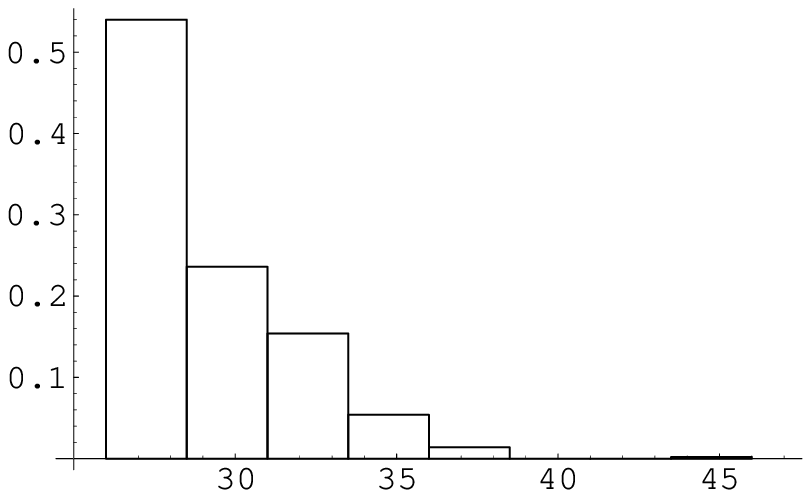}\\
\hspace{0.5cm}\parbox[t]{3.5cm}{c) $k_S=2$, $r=1$.}
\hspace{0.5cm}\parbox[t]{3.5cm}{d) $k_S=2$, $r=17$.}\\
\includegraphics[scale=0.453]{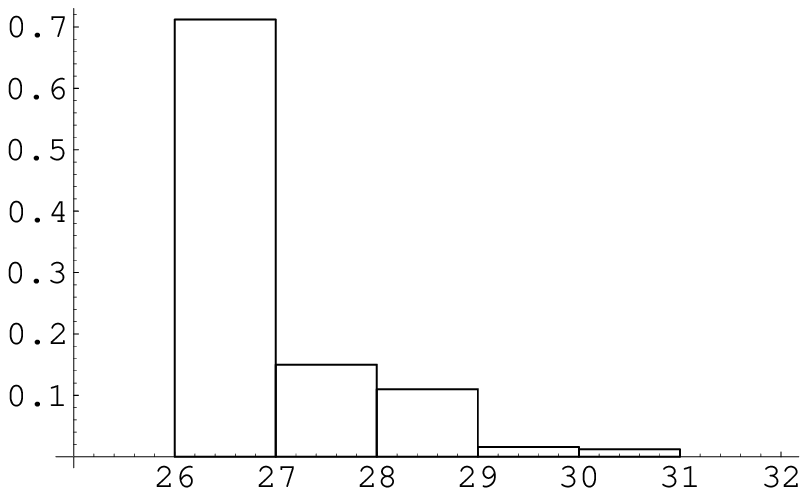}
\includegraphics[scale=0.453]{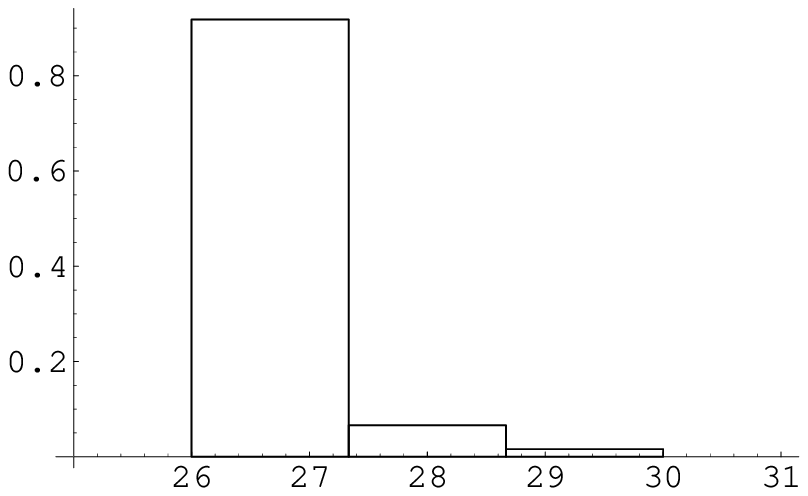}\\
\hspace{0.5cm}\parbox[t]{3.5cm}{e) $k_S=4$, $r=1$.}
\hspace{0.5cm}\parbox[t]{3.5cm}{f) $k_S=4$, $r=17$.}\\
\caption{Total evacuation time distribution for different $k_S$
and $r$ over 500 experiments.}\label{gist}
\end{center}
\end{figure}

Figures \ref{gist} show total evacuation time distributions for
some couples of the parameters from table~\ref{Tmo} over 500
realizations.

\begin{figure}[!h]
\begin{center}
\includegraphics[scale=0.331]{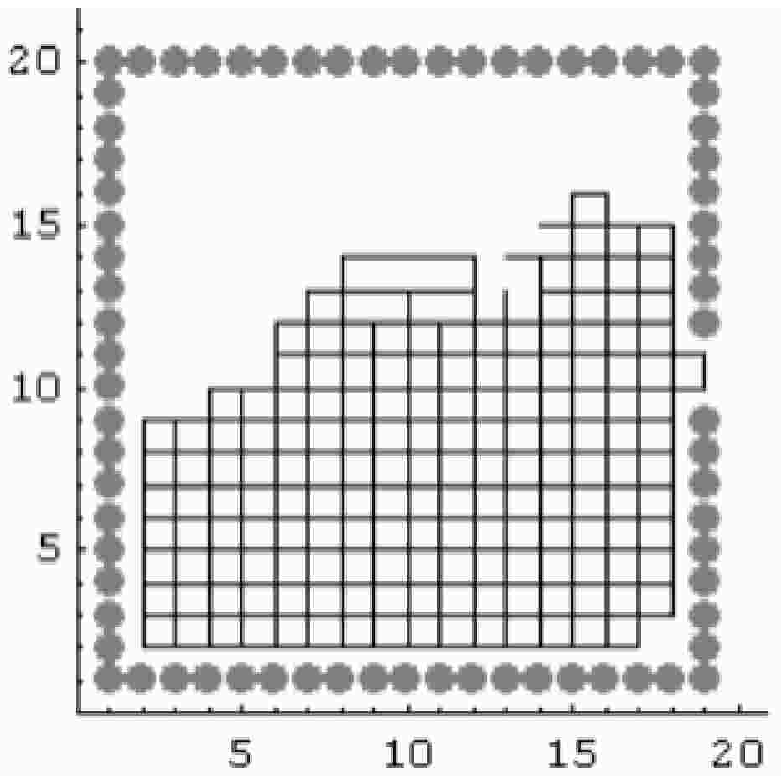}
\includegraphics[scale=0.331]{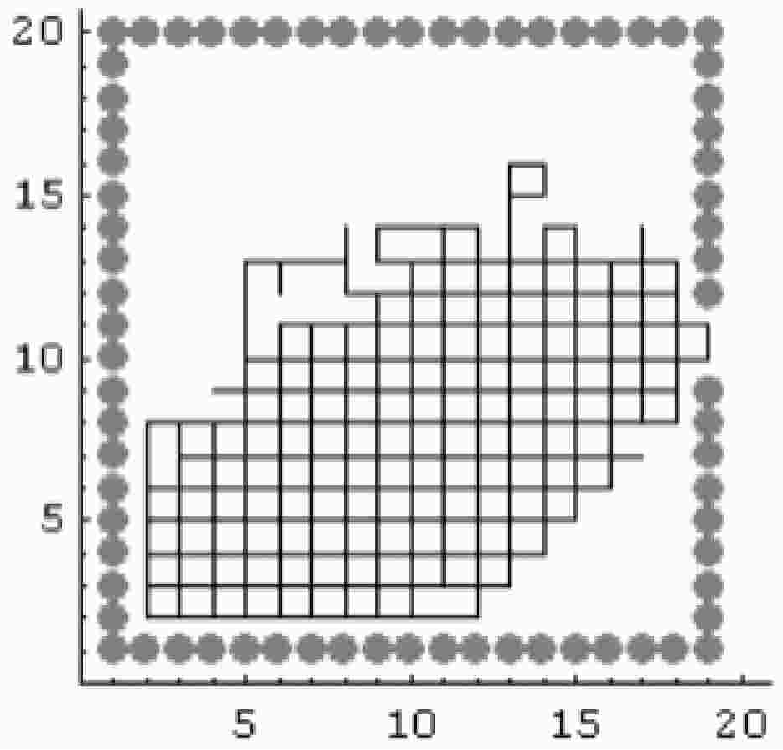}\\
\hspace{0.5cm}\parbox[t]{4cm}{a) $k_S=1$, $r=1$.}
\hspace{0.5cm}\parbox[t]{3cm}{b) $k_S=1$, $r=17$.}\\
\includegraphics[scale=0.331]{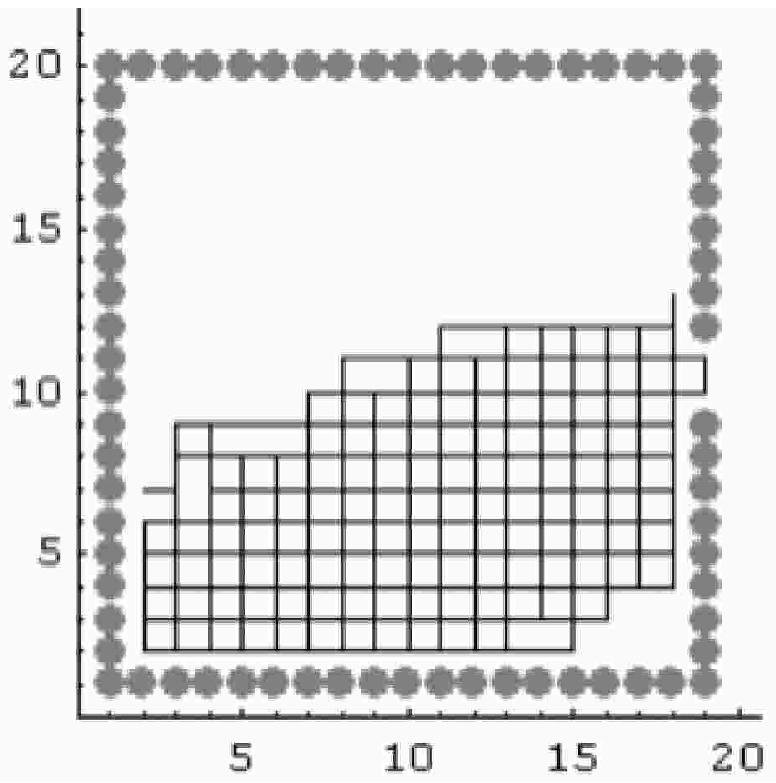}
\includegraphics[scale=0.331]{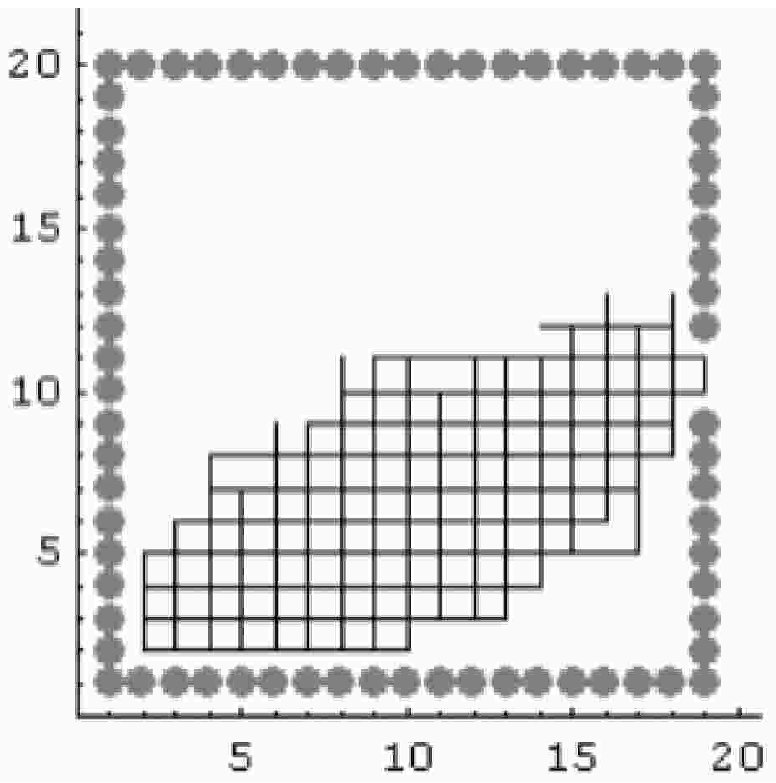}\\
\hspace{0.5cm}\parbox[t]{4cm}{c) $k_S=2$, $r=1$.}
\hspace{0.5cm}\parbox[t]{3cm}{d) $k_S=2$, $r=17$.}\\
\includegraphics[scale=0.331]{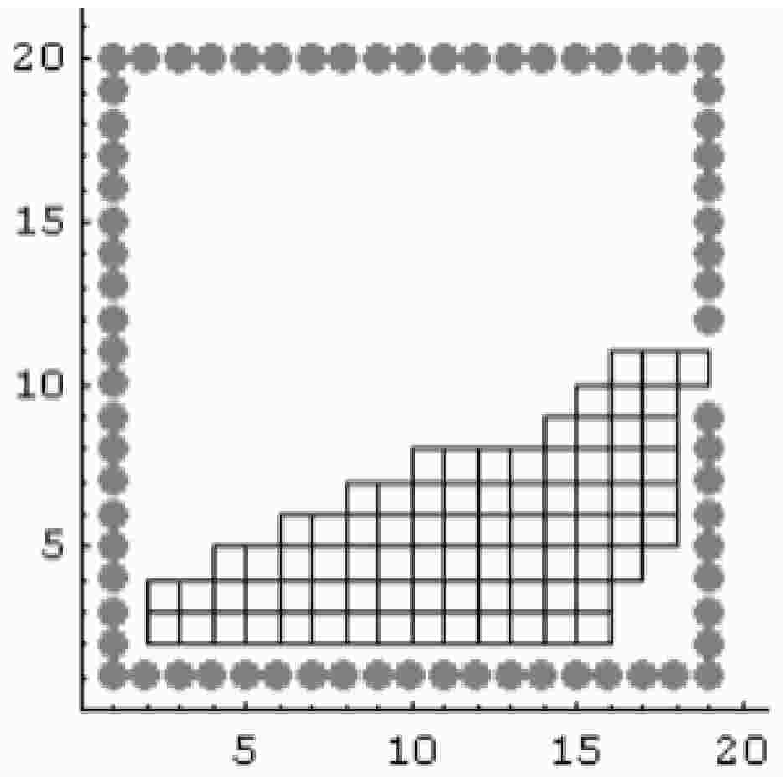}
\includegraphics[scale=0.331]{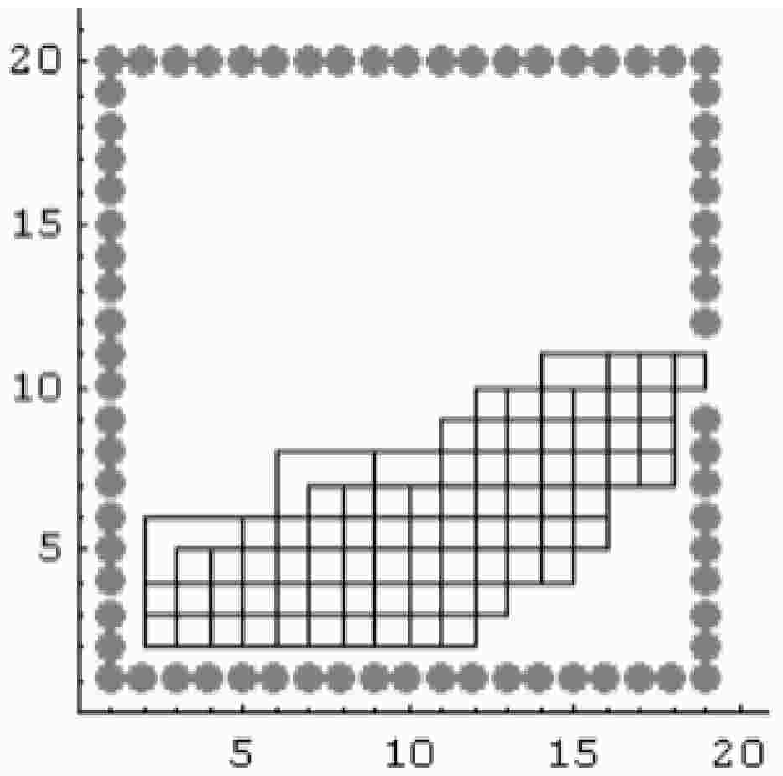}\\
\hspace{0.5cm}\parbox[t]{4cm}{e) $k_S=4$, $r=1$.}
\hspace{0.5cm}\parbox[t]{3cm}{f) $k_S=4$, $r=17$.}\\
\caption{Tracks of one pedestrian for different $k_S$ and $r$ over
500 experiments.}\label{treks}
\end{center}
\end{figure}

Figures \ref{treks} present tracks of pedestrian ways over 500
realizations for some couples of the parameters from
table~\ref{Tmo}.

Different moving conditions are reproduced by these combinations.
They vary from  $r=1$, $k_S=1$ to $r=17$, $k_S=4$. The former case
can be interpreted as pedestrian moves in a very low visibility
(or by touch) but he approximately knows the direction to
destination point. In other words one can say that pedestrian
doesn't see, knows, and wants not so much to go to destination
point (exit). Last one ($r=17$, $k_S=4$) describes situation when
pedestrian sees, knows, and wants to go to destination point very
much.

Note that if $r=1$ model presented corresponds to
FF-model~\cite{SimPedDynFFM, SimEvacuatFFM} with the same other
parameters. And it's clear that the pedestrian patience that was
introduced in the model doesn't pronounced in one pedestrian case.

One can see that for small $k_S=1$ mode $T_{mo}$ is very dependent
on parameter $r$. The bigger parameter $k_S$ is an influence of
$r$ to $T_{mo}$ is less pronounced. But the bigger parameter $r$
is more natural way pedestrian chooses --- tracks are more close
to a line connecting starting point and exit (non the less random
component takes place). Under $r>1$ 
a proximity of wall decreases the probability (\ref{1}) to move in
this direction. Thereby tracks are forced to tend to natural one.
Thus parameter $r$ fulfils its role to simulate environment
analysis here.

\subsection{Many pedestrians cases}
\subsubsection{Influence of $k_S$ and $r$}
\begin{figure}
\begin{center}
\hspace{0.5cm}\parbox[t]{2cm}{$t=25$}
\hspace{0.5cm}\parbox[t]{2cm}{$t=185$}
\hspace{0.5cm}\parbox[t]{2cm}{$t=470$}\\
\includegraphics[scale=0.3]{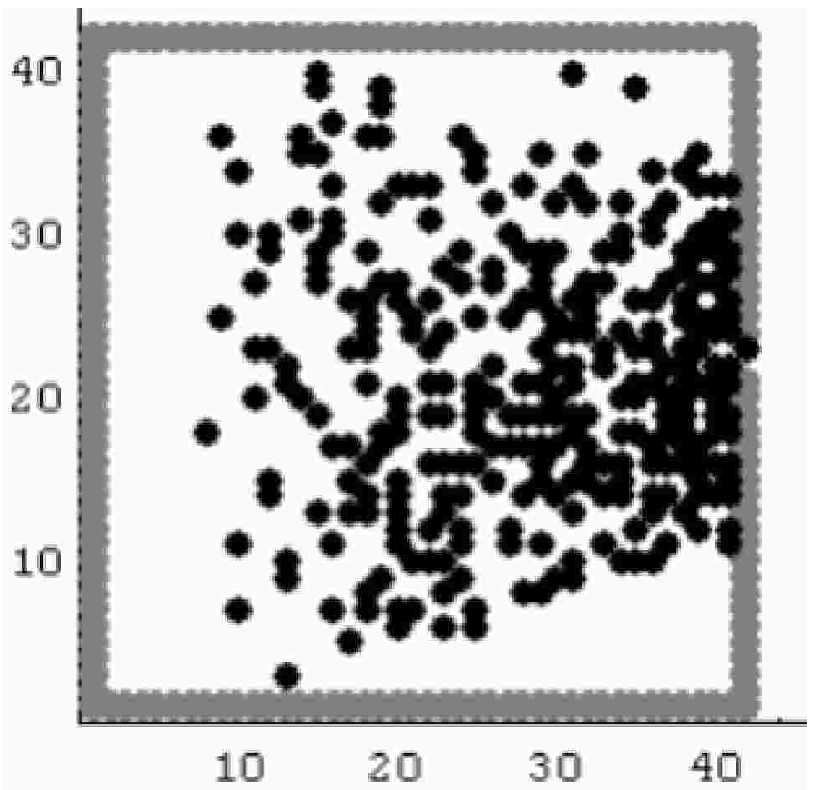}
\includegraphics[scale=0.3]{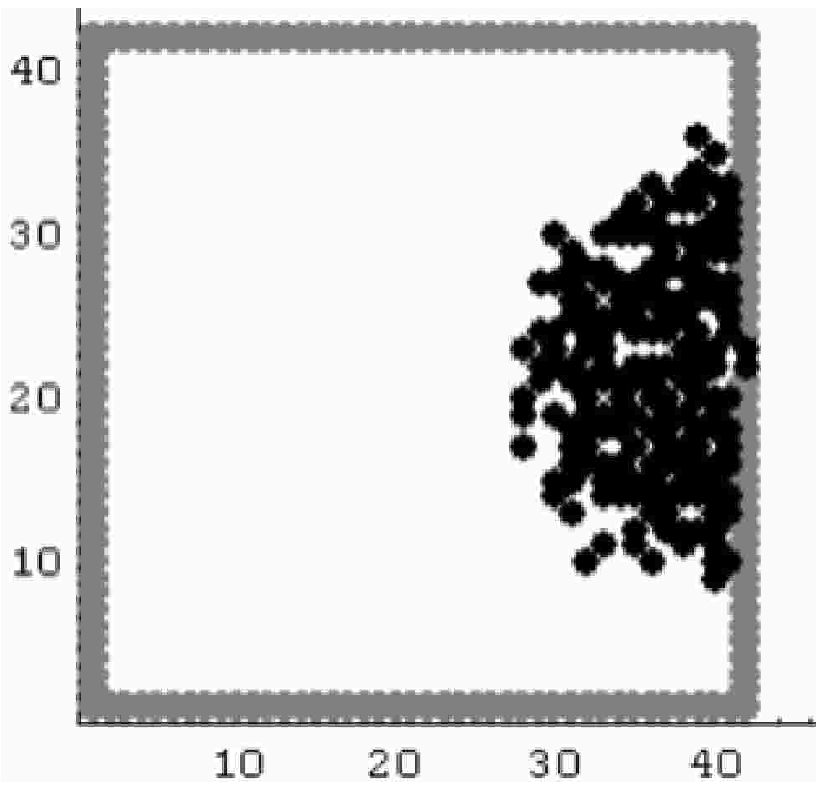}
\includegraphics[scale=0.3]{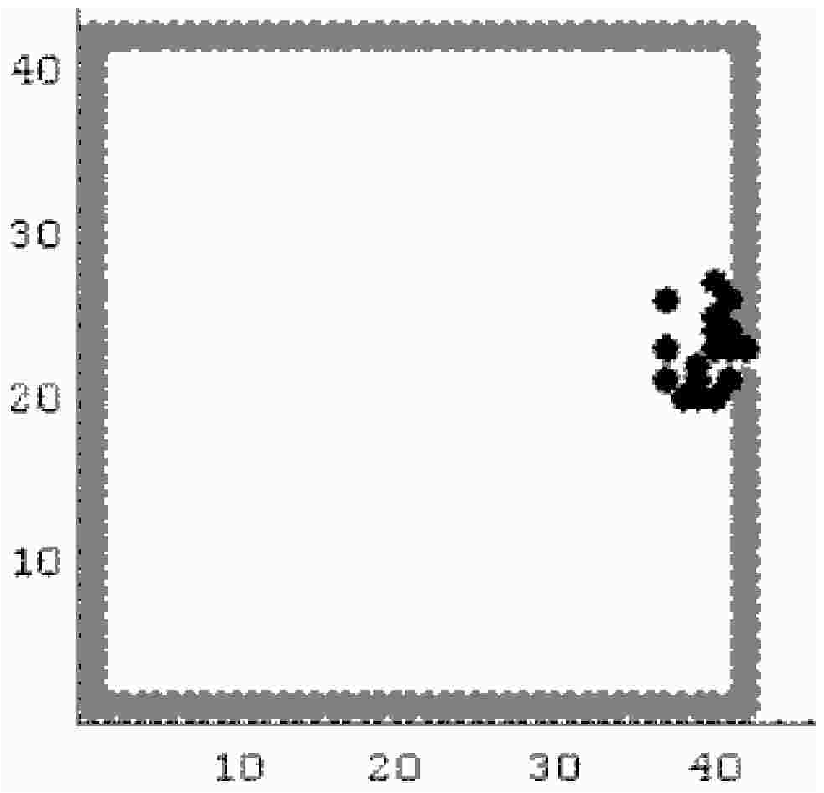}\\
\hspace{0cm}\parbox[t]{8cm}{a) $k_S=1$, $r=1$, $T_{total}=509$}\\[10pt]
\hspace{0.5cm}\parbox[t]{2cm}{$t=20$}
\hspace{0.5cm}\parbox[t]{2cm}{$t=210$}
\hspace{0.5cm}\parbox[t]{2cm}{$t=500$}\\
\includegraphics[scale=0.3]{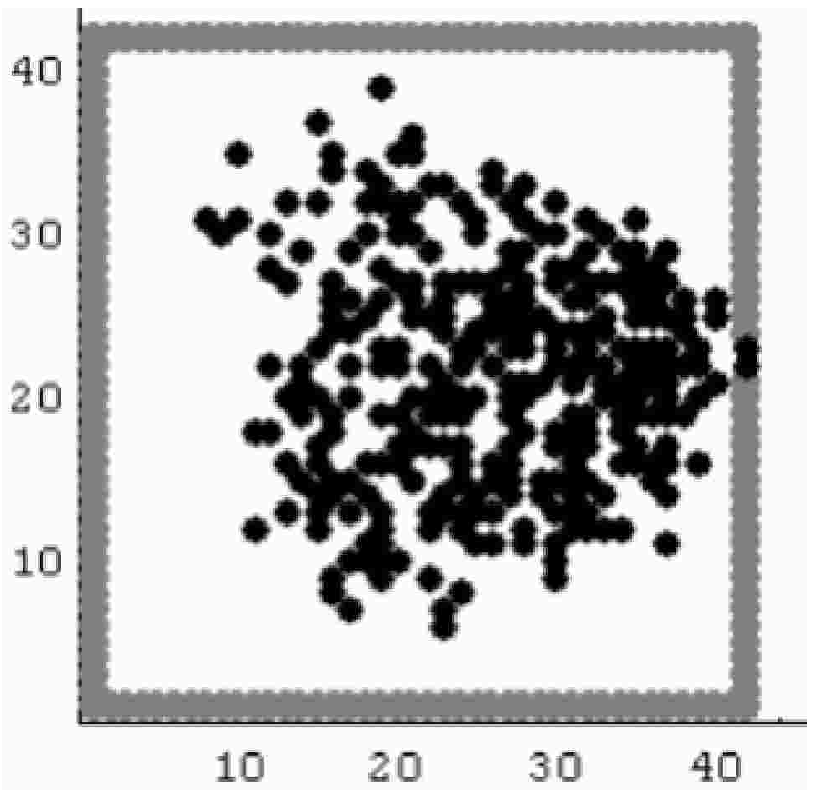}
\includegraphics[scale=0.3]{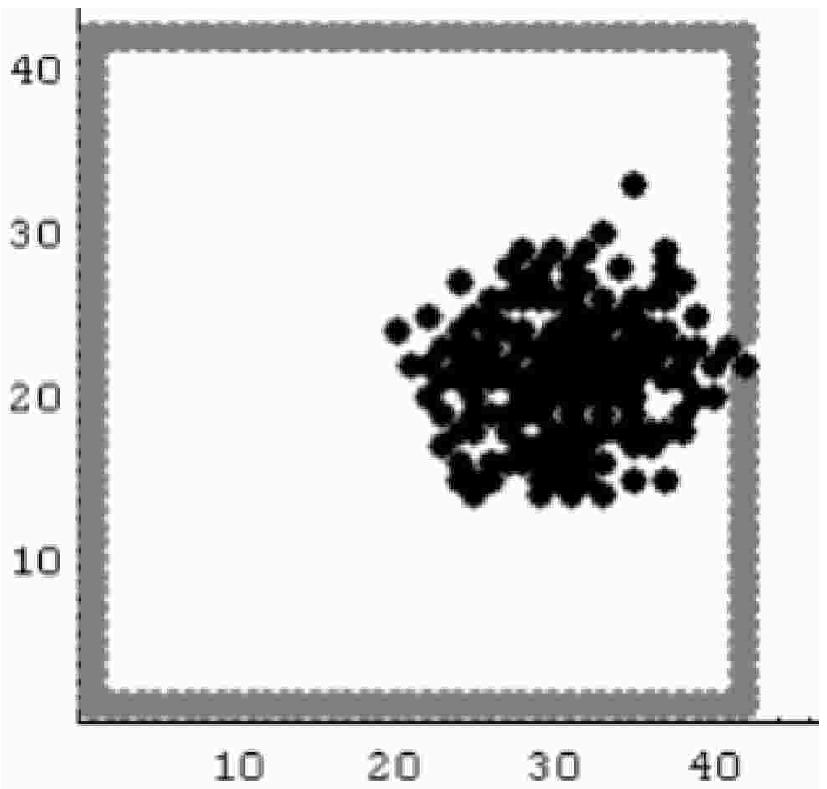}
\includegraphics[scale=0.3]{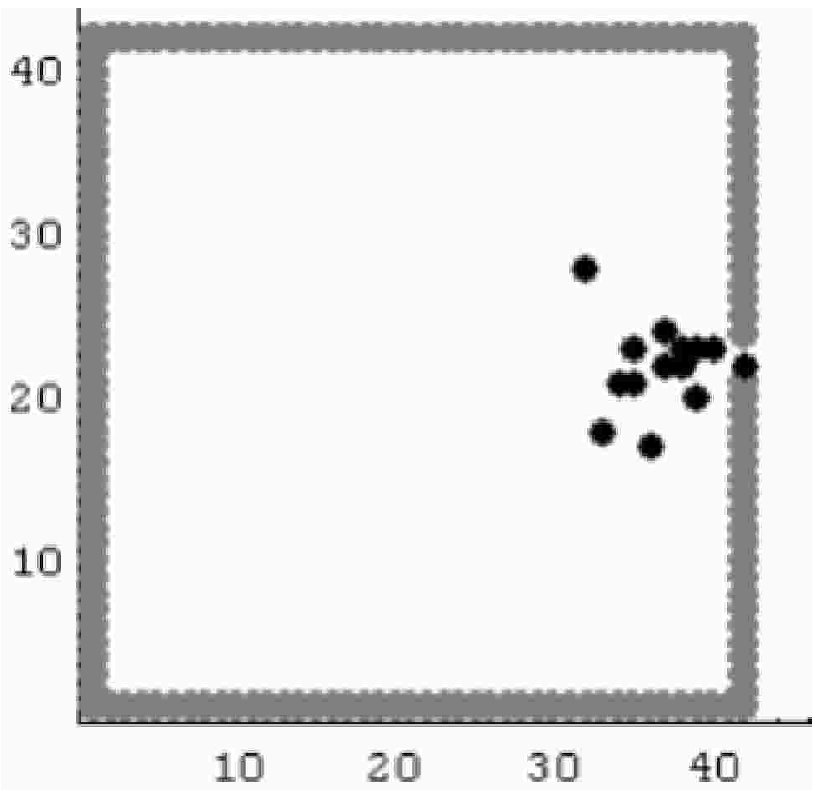}\\
\hspace{0cm}\parbox[t]{8cm}{b) $k_S=1$, $r=40$, $T_{total}=603$}
\caption{Evacuation stages for 300 people for $k_S=1$ and
different $r$.} \label{300people_ks1}
\end{center}
\end{figure}

\begin{figure}
\begin{center}
\hspace{0.5cm}\parbox[t]{2cm}{$t=15$}
\hspace{0.5cm}\parbox[t]{2cm}{$t=100$}
\hspace{0.5cm}\parbox[t]{2cm}{$t=310$}\\
\includegraphics[scale=0.3]{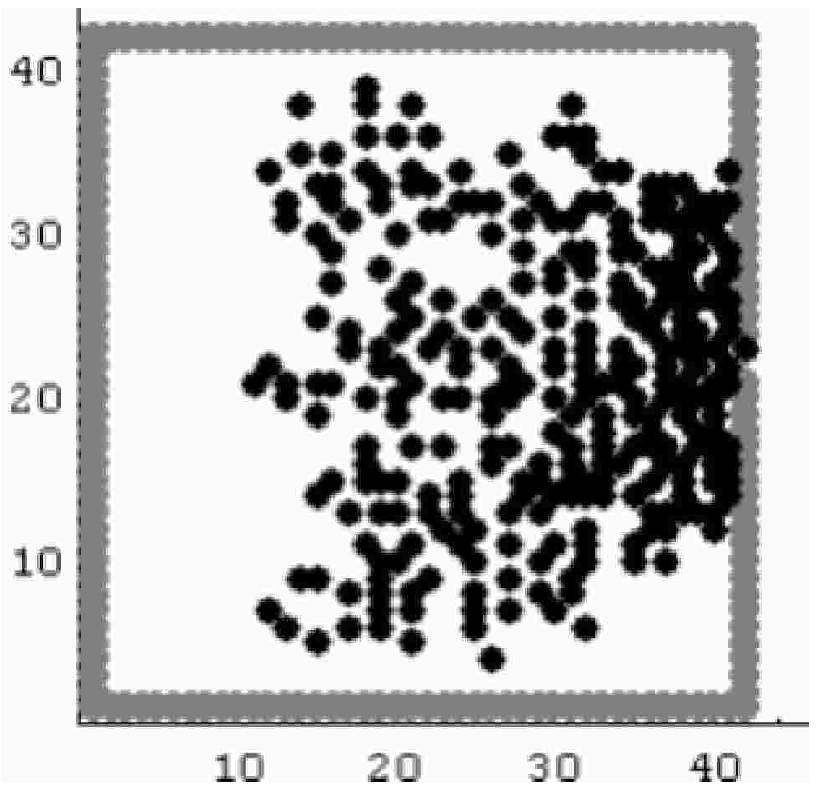}
\includegraphics[scale=0.3]{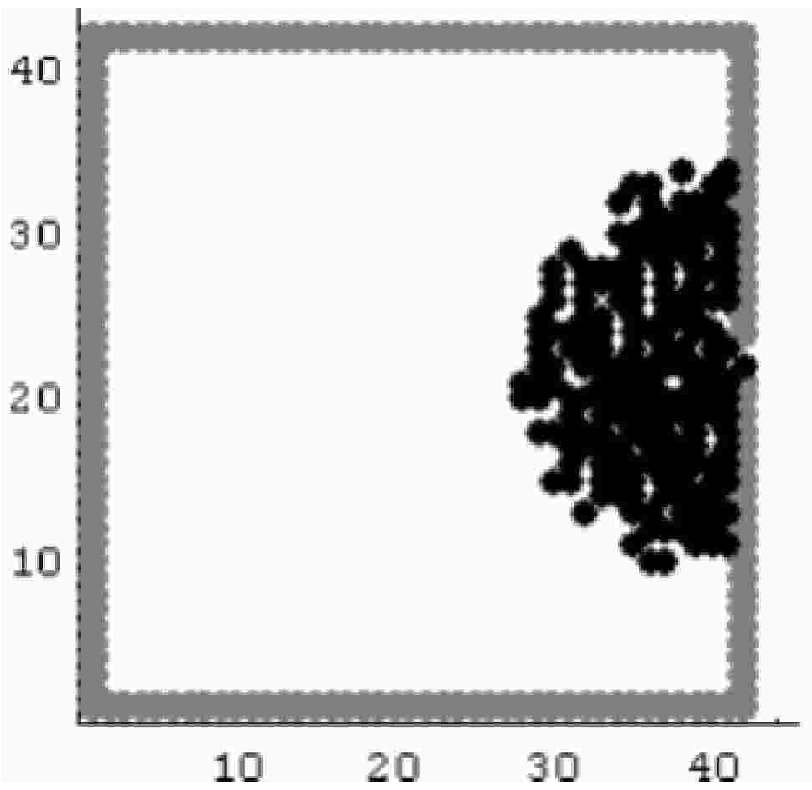}
\includegraphics[scale=0.3]{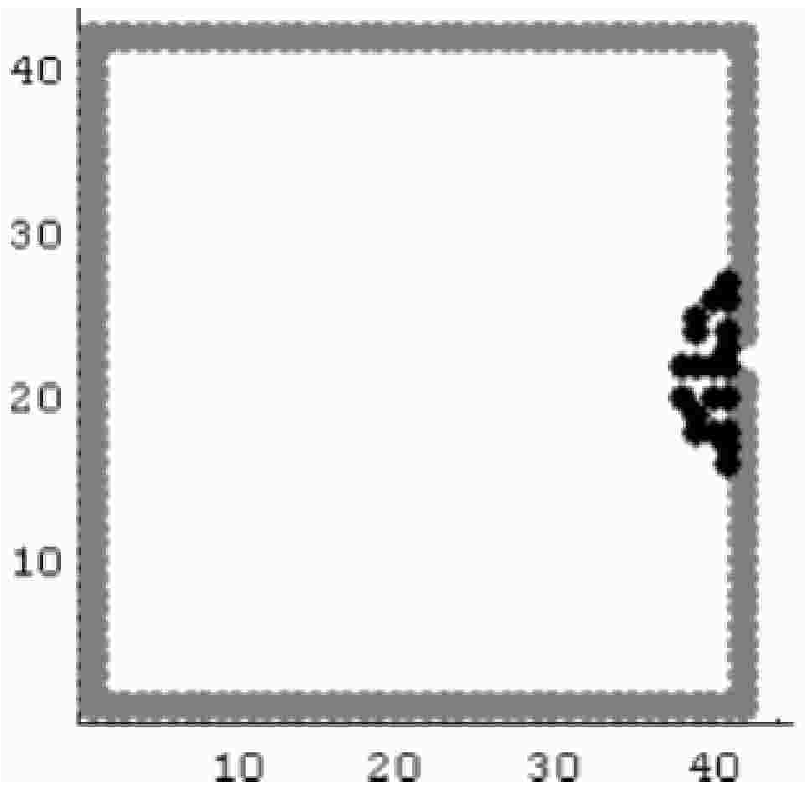}\\
\hspace{0cm}\parbox[t]{8cm}{a) $k_S=3$, $r=1$, $T_{total}=336$}\\[10pt]
\hspace{0.5cm}\parbox[t]{2cm}{$t=15$}
\hspace{0.5cm}\parbox[t]{2cm}{$t=80$}
\hspace{0.5cm}\parbox[t]{2cm}{$t=285$}\\
\includegraphics[scale=0.3]{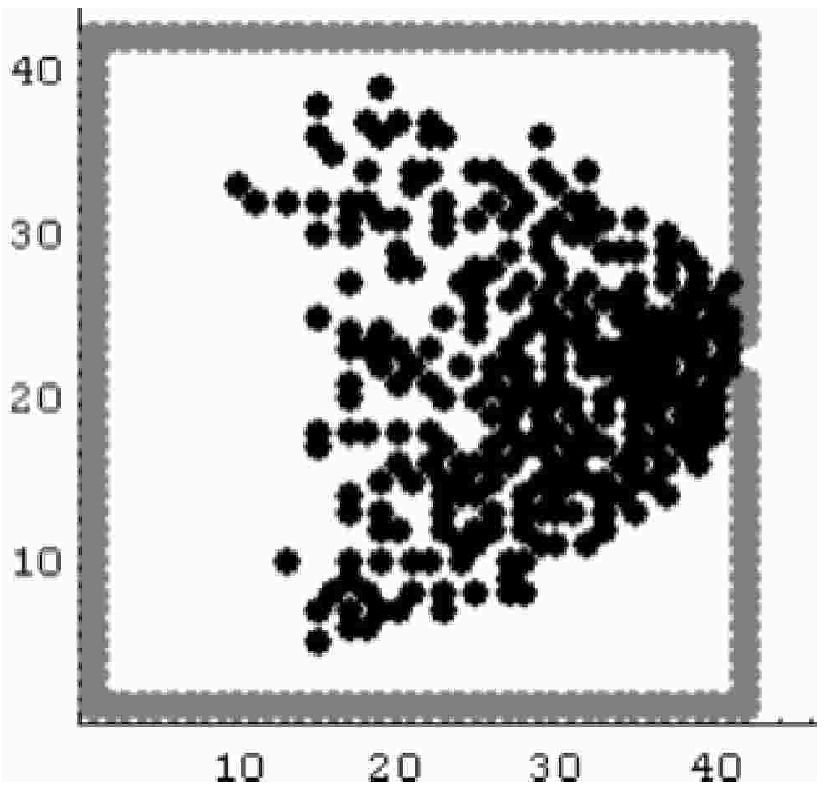}
\includegraphics[scale=0.3]{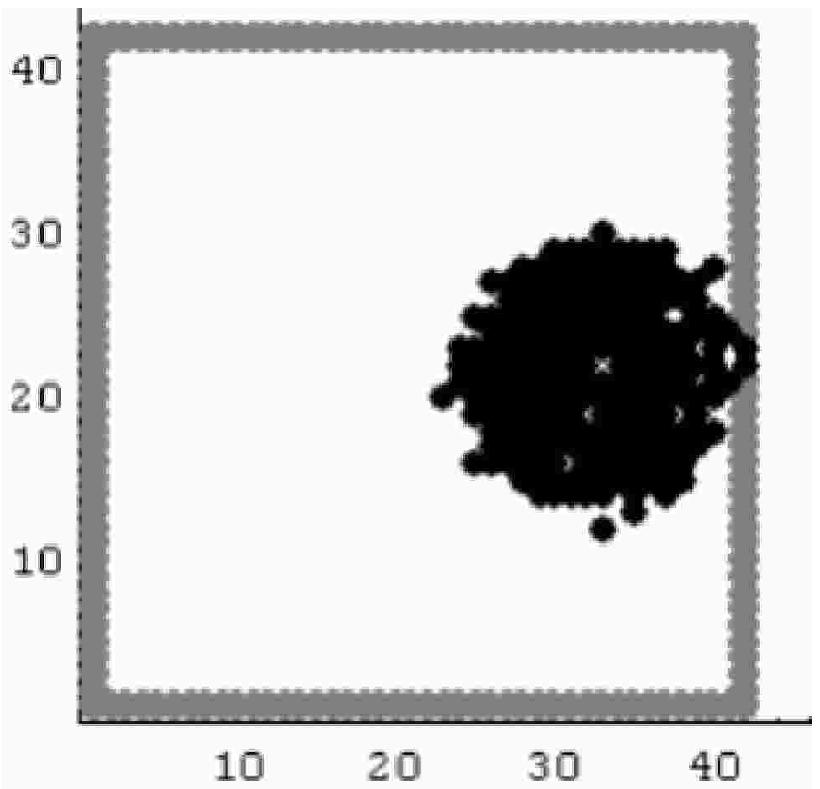}
\includegraphics[scale=0.3]{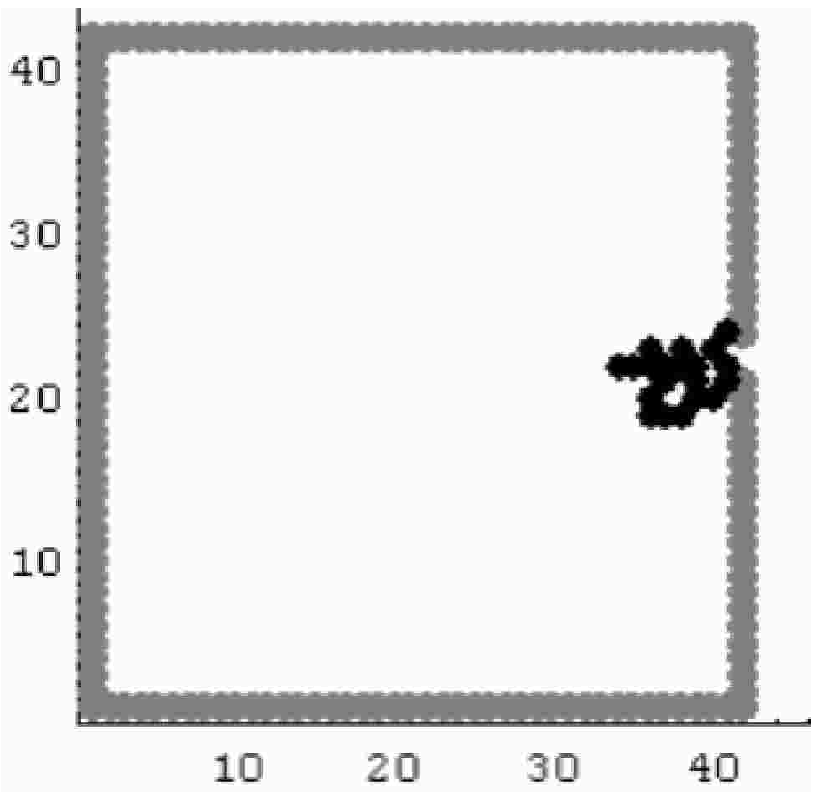}\\
\hspace{0cm}\parbox[t]{8cm}{b) $k_S=3$, $r=40$, $T_{total}=317$}
\caption{Evacuation stages for 300 people for $k_S=3$ and
different $r$.} \label{300people_ks3}
\end{center}
\end{figure}

For this collective experiment the space was a room $16m\times16m$
($40$ cells $\times$ $40$ cells) with one exit ($0.8m$). Initial
number of people is $N=300$ (density $\rho\approx0.19$). Initial
positions are random and people start to move towards the exit
with $v=v_{max}=1$. Exit is in the middle of east wall. Figures
\ref{300people_ks1}, \ref{300people_ks3} present typical stages of
evacuation process for different $k_S$ and $r$.

One can see that evacuation dynamics in case a) differs from case
b) in both figures \ref{300people_ks1}, \ref{300people_ks3}. The
reason of it is different parameters $r$. If $r>1$ people avoid to
approach to walls. And while crowd density allows pedestrians try
to follow more natural way to the exit.  The bigger $r$ is shape
of crowd in front of exit is more diverse from the case of $r=1$.
(Note if $r=1$ we have FF-model with the same other parameters.)

One can notice that the closer to the exit pedestrians are circle
shaped crowd in front of the exit is more unrealistic. The problem
comes from computational aspect of coefficient $A_{ij}$. It's a
positive that in the model obtained people avoid to approach to
walls. But this effect has to be less pronounced with approaching
to wall (walls as in a subsection example below) surrounding exit.
Thus parameter $r$ has to be spatial adaptive.

Next table demonstrates numerical description of cases presented.
Let $f_{\alpha}$ be frequency to choose direction
$\alpha=\{N,E,S,W,C\}$ over all experiment for each couple of
parameters. Here $N,E,S,W,C$ are north, east, sought, west, center
(stay at present place) correspondingly, $M$ --- total number of
movements including stayings at current position over all
experiment.

\begin{table}[!h]
\caption{Direction frequencies.}\label{distrDirect}
\fbox{
\begin{tabular}{lrrrrrr} \,\,\,\,\,$k_S$, $r$ &
$f_N$ & \multicolumn{1}{c}{$f_S$} & \multicolumn{1}{c}{$f_W$}&
\multicolumn{1}{c}{$f_E$} & \multicolumn{1}{c}{ $f_C$}&
\multicolumn{1}{c}{ $M$} \\[4pt]
1) $1$, $1$ & 0.23 & 0.23 & 0.17 & 0.27 & 0.08 & 77961\\
2) $1$, $40$ & 0.16 & 0.16 & 0.10 & 0.20 & 0.38 & 77976\\
3) $3$, $1$ & 0.21 & 0.2 & 0.13 & 0.31 & 0.15 & 49313\\
4) $3$, $40$ & 0.06 & 0.06 & 0.01 & 0.18 & 0.69 & 47133
\end{tabular}}
\end{table}

Comparing cases 1) and 2), 3) and 4) correspondingly one can
notice that greater $r$ leads to significant redistribution of
flow. In the case of $r=1$ low $f_C$ says that people move as much
as possible (because people can stay at current position if all
nearest cells are occupied only). If $r=40$ the opportunity to
wait is realized --- in cases 2) and 4) $f_C$ has the greatest
value. $f_W$ has the smallest value (this direction is opposite to
the  exit), $f_N$, $f_S$ are approximately equal because exit is
in the middle of the wall. Thus in cases 2) and 4) model
reproduces more natural decision-making process.

At the same time let us remark that increasing $k_S$ makes
evacuation process more directed and reduces total evacuation time
(see cases 1) and 3), 2) and 4) correspondingly).

\subsubsection{Influence of exit position}
\begin{figure}[!h]
\begin{center}
\includegraphics[scale=0.453]{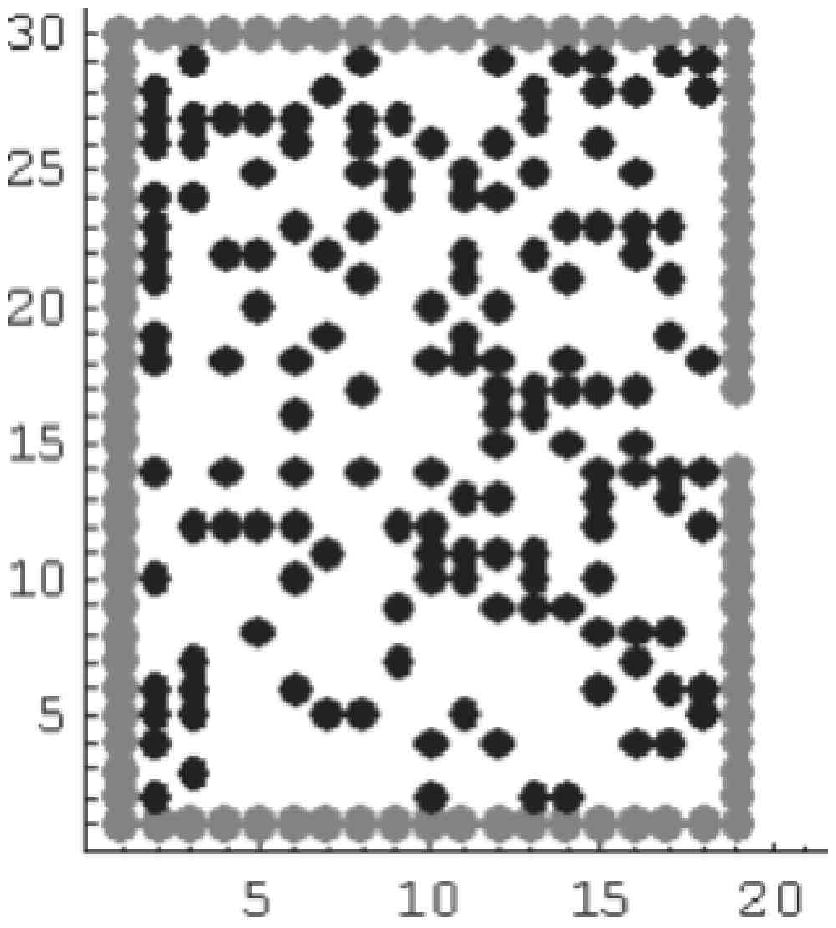}
\includegraphics[scale=0.453]{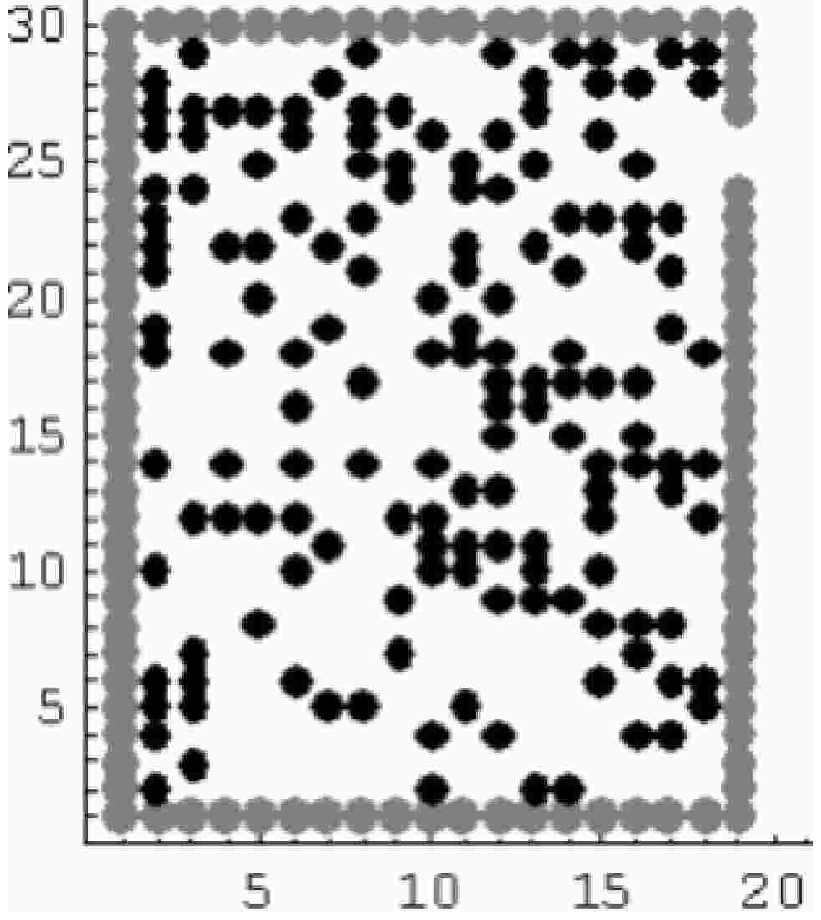}\\
\hspace{0.5cm}\parbox[t]{3.5cm}{a)}
\hspace{0.5cm}\parbox[t]{3.5cm}{b)}\\
\caption{Initial people position and exits
position.}\label{150people}
\end{center}
\end{figure}

For this collective experiment the space was a room
$6.8m\times11.2m$ ($17$ cells $\times$ $28$ cells) with one exit
($0.8m$). Initial $N=150$ (density $\rho\approx0.31$). Initial
positions are random and people start to move towards the exit
with $v=v_{max}=1$. Spaces are presented in the figures
\ref{150people}a, \ref{150people}b.

Two combinations of parameters $k_S$ and $r$ were considered. And
total evacuation time was investigated. Table \ref{TmoExits}
contains results over 100 experiments. Initial positions of people
were the same for all experiments.

\begin{table}[!h]
\center \caption{Modes $T_{mo}$ of total evacuation time
distributions with middle exit (a) and corner exit (b).}
\label{TmoExits} a)\hspace{0.2cm} \fbox{
\begin{tabular}{lrr} $k_S$ & \multicolumn{1}{c}{$r$} &
\multicolumn{1}{c}{$T_{mo}$}\\[4pt]
3 & 2 & 158 \\
3 & 20 & 160
\end{tabular}}
\hspace{0.5cm}
b)\hspace{0.2cm} \fbox{
\begin{tabular}{lrr} $k_S$ & \multicolumn{1}{c}{$r$} &
\multicolumn{1}{c}{$T_{mo}$}\\[4pt]
3 & 2 & 174 \\
3 & 20 & 226
\end{tabular}}
\end{table}

One can notice  once again that in the case a) parameter $r$
doesn't influence on total evacuation time under such $k_S$. But
in other case increase of $r$ leads to significant delay of
evacuation. The reason of it is ``computational'' repulsion from
the walls. And as a result pedestrians don't use corner between
wall and exit, exit is not fully used, total evacuation time
increases. So this is one more example that shows necessity in at
least  parameter $r$ adaptivity.

\section{Conclusion and further plans}
In the paper the intelligent FF cellular automation model is
presented. Modifications made are to improve realism of the
individual pedestrian movement simulation. The following features
of people behavior are introduced: keeping apart from other people
(and obstacle), patience.
 Idea of spatial adaptation
of model parameters is pronounced  and one method is presented
(parameter $\tilde \mu_{ij}$). Model obtained saved opportunities
to reproduce variety of collective effects of pedestrian movement
from free walk to escape panic and took more flexibility.

The simulation made showed real improvements in decision-making
process in comparison with basic FF model and pointed out some
problems. The following points seem to be very important for
realistic pedestrian simulation and are under future
investigations.

\subsection{Jam}
In a case of emergency appearing of clogging situation in front of
exit often leads to appearing fallen or injured people (or jam).
The physical interactions in such crowd add up and cause dangerous
pressures up to $4,450Nm^{-1}$ ~\cite{HelbingOV} which ``can bend
steel barriers or push down brick walls''. Fallen or injured
people act as ``obstacles'', and escape is further slowed.
Continuous model can reproduce pushing and physical interactions
among pedestrians. CA model doesn't allow to do it. Parameter
$\tilde \mu_{ij}$ works in the model as some kind of local
pressure between the pedestrians (the higher $\tilde \mu_{ij}$ is
pedestrians are more handicapped by others trying to reach the
same target cell). But by means of $\tilde \mu_{ij}$ fallen or
injured people are not simulated. Our further intention is to
produce method to evaluate common pressure to each pedestrian in
CA model, and if pressure is over some barrier to indicate
correspondent cell as new obstacle.

\subsection{Parameters adaptation}
There are at least two reasons for parameters to be adaptive. One
of them is learning that is reside to people. Therefor at least
parameters $k_S$ and $k_D$ need to be time adaptive and spatial
dependent as well. Parameter $r$ needs to be spatial adaptive
because of negative computational effects. So methods to adapt
model parameters are under further investigation.

\small {\subsubsection{Corresponding author's biography.}
Dr.Ekaterina Kirik is  a scientific research fellow at the
Institute of Computational Modelling of Siberian Branch of Russian
Academy of Sciences, teacher at the Siberian Federal University.
She has got PhD degree from 2002. She published about 35 papers
and thesis, took part in many international conferences with oral
and poster presentations. Research interests are data analysis,
data mining, robust estimating, pattern recognition, statistical
modelling, pedestrian behavior modelling, nonparametric statistic.
She is experienced in some industrial applications and
international collaboration.}
\end{document}